# Food, Affection and Gaze: Which Cues do Free-Ranging Dogs Consider for Engaging with Humans?

Srijaya Nandi[a], Dipayanti Aditya[b], Tithi Chakraborty[a], Rachael Sara Paul[c], Anindita Bhadra[a*]

[a]Department of Biological Sciences, Indian Institute of Science Education and Research, Kolkata, Mohanpur, Nadia, West Bengal, India

[b]Department of Biotechnology, Maulana Abul Kalam Azad University of Technology, Haringhata, Nadia, West Bengal, India

[c]National Institute of Science Education and Research, Bhubaneswar, Khurda, Odisha, India

Corresponding Author:

Anindita Bhadra[a*]

Email address: abhadra@iiserkol.ac.in

**Abstract**

Free-ranging dogs (*Canis lupus familiaris*) constitute the majority of the global dog population and rely heavily on human-derived resources. Studies show different levels of responses to various cues like food, petting and gazing by humans. However, the relative importance that dogs associate with these rewards, driving their interactions with unfamiliar humans remain understudied. Understanding how these dogs prioritize different rewards, ranging from food to social contact, can offer insights into their adaptive strategies within human-dominated ecosystems, and help to reduce conflict. We investigated the motivational value of different reward types in 150 adult free-ranging dogs in West Bengal, India. Using a between-subjects design, unfamiliar experimenters offered one of five rewards: high-value food (chicken), low-value food (biscuit), social interaction (petting), human gaze only, or human presence only. Motivation was assessed by measuring the number of rewards accepted, approach latency, duration of proximity, and behaviour via a Socialization Index (SI). High-value food was the most potent driver of approach behaviour and sustained proximity. While petting elicited higher SI scores, indicating affiliative engagement, it was associated with more rapid satiation than food. Human gaze alone functioned as a subtle reinforcer compared to passive presence, maintaining dog attention longer than presence alone. These findings suggest that free-ranging dogs prioritize high-energy intake over social interaction with strangers, consistent with an optimal foraging strategy. This behavioural flexibility enables them to balance energy needs against potential risks, demonstrating the sophisticated decision-making crucial for survival in urban environments where humans act as both resource providers and potential threats.

**Introduction**

Dogs (*Canis lupus familiaris*) have shared a unique and enduring bond with humans since time immemorial. This bond is often compared to that observed in mother-infant relationships (Serpell, 2003). Dogs are also frequently referred to as man's best friend (Phillips, 1975). Unlike other domesticated animals, which were primarily domesticated for purposes such as food, clothing or transportation, the purpose behind dog domestication remains uncertain (Larson et al., 2012; Zhang et al., 2020). It is believed that domestication took place in various parts of Europe and/or Asia between 15,000 and 40,000 years ago (Larson et al., 2012; Ovodov et al., 2012; Frantz et al., 2016). Several theories attempt to explain dog domestication, the most prominent of which are the self-domestication and cross-species adoption hypotheses (Serpell, 2021).

Social attachment with humans is also believed to have played a key role in domestication (Nagasawa et al., 2015). Given the long-shared history of close interactions between dogs and humans, it is intriguing to explore the reinforcers that drive and maintain this relationship. The most significant among these reinforcers are petting, food, and vocal praise. Multiple studies on owned or pet dogs have addressed this issue, with some demonstrating a greater relative importance of food compared to petting, and others reporting the contrary (Fukuzawa and Hayashi, 2013; Feuerbacher and Wynne, 2012; Elliot and King, 1960; Brodbeck, 1954; Igel and Calvin, 1960; Fonberg et al., 1981; Bhattacharjee et al., 2017).

The duration of social interaction may also influence its effectiveness relative to food. Fonberg et al. (1981) found that 20-30 seconds of petting was as effective as food in maintaining behaviour. However, food proved more effective than petting when the duration of interaction was limited to 4 seconds (Feuerbacher & Wynne, 2012). Some studies also suggest that the mere presence of a human can act as a primary reinforcer for dogs. For example, puppies were observed to run faster toward an endbox when a passive person was present, compared to when it was empty (Stanley et al., 1965).

A study on free-ranging dogs showed that petting can be more effective than a food reward over repeated interactions for building trust (Bhattacharjee et al., 2021). Though the free-ranging dogs make up approximately 80% of the global dog population (Boitani and Ciucci, 1995; Hughes and Macdonald, 2013), in-depth studies testing the impact of different reward types on their interactions with unfamiliar humans are scarce. These dogs live in close proximity to humans but are not under direct human supervision or control (Boitani et al., 2017). Unlike pet dogs, they don't have an owner to take care of them. They live independently on the streets around human habitations and are found ubiquitously along the rural-urban continuum (Berman & Dunbar, 1983). Pariah dogs living in parts of Asia are considered to be living relics of the early associations that existed between primitive humans and wild canids (Downs, 1960; Zeuner, 1963). The free-ranging dogs are heavily reliant on humans for food and often give birth near human settlements (Sen Majumder et al., 2016). The relationship between humans and free-ranging dogs is complex, with public attitudes ranging from affection to hostility (Bhattacharjee et al., 2020). On the one hand, some people provide active care, shelter and food to these dogs, while some consider them a menace. Humans are the largest single factor for the high early-life mortality of the free-ranging dogs in India (Paul et al., 2016). Cases of dog-human conflict are on the rise in urban areas of India (Rai, 2025), which makes understanding the sources of conflict and means of co-existence imperative for the well-being of both dogs and humans. Understanding the factors that shape the social behaviour of free-ranging dogs toward humans can provide valuable insights into the evolutionary dynamics of domestication. The present study focuses on this population of dogs from the Indian subcontinent.

Adult free-ranging dogs in India have been shown to exhibit a clear preference for meat over other types of food (Bhadra and Bhadra, 2014). Given their reliance on human-generated food sources, food is widely regarded as a primary reinforcer for these dogs. This dependency not only influences their foraging strategies but also shapes their interactions with humans. However, there remains a significant gap in the literature regarding whether the quality or type of food reward, for example, meat versus biscuits, affects free-ranging dogs' social behaviour and proximity-seeking towards the individual providing the food. Understanding whether different food types modulate free-ranging dogs' approach behaviour could offer deeper insights into their motivation and learning processes.

Different parameters can be used to measure the potential of a reward. Approach latency, defined as the time taken by a subject to approach a given stimulus, is commonly used as an indicator of the motivational value of different rewards (Feuerbacher et al., 2012). Another relevant behavioural measure is the amount of time an animal spends in proximity to a human, which has been widely used as an indicator of sociability in non-human animals (Barrera et al., 2010). Furthermore, in attachment theory, proximity seeking and maintenance are considered core behavioural expressions of attachment bonds in humans (Bowlby, 1973), and this framework has been extended to assess human-animal relationships as well. Together, these behavioural metrics provide valuable insights into the strength and nature of affiliative and motivational processes underlying dog-human interactions.

Another important aspect of reinforcement is the satiation potential of the reward, that is, how quickly its effectiveness diminishes with repeated exposure. For instance, puppies (of pet dogs) given access to a passive human before a trial were observed to run more slowly than those denied such access, suggesting that the reinforcing value of human presence may decrease following prior exposure (Bacon & Stanley, 1963). Food has been shown to have lower satiation effects than petting, meaning it tends to retain its reinforcing value across repeated trials, thereby functioning as a more consistent and effective motivator in operant tasks (Feuerbacher & Wynne, 2012).

A study conducted on free-ranging dogs in Morocco showed that the dogs spent more time in proximity to the cuddle provider compared to the food provider, pointing to the hypersociability hypothesis (Lazzarroni et al., 2020). However, when given a choice between them, the choice was based on chance. An earlier study free-ranging dogs in India showed that the dogs prefer food over petting during a one-time interaction, but show a preference for both rewards during repeated interactions (Nandi et al, 2025). However, this study only used single presentations of rewards and did not examine how dog behaviour changes when rewards are given multiple times, or are of differing qualities, in one interaction.

Understanding how reward types influence dog behaviour, particularly in free-ranging populations, can help understand the behavioural adaptations involved in domestication. Identifying the relative efficacy of food versus social rewards in maintaining or initiating social interactions is key to grasping how human-dog relationships evolved and persisted in diverse ecological contexts.

We aimed to investigate the relative importance of various reward types on the motivation of free-ranging dogs: human presence alone, human presence with gazing, petting combined with gazing, and gazing accompanied by food provisioning (either biscuits or chicken). We were also interested to understand whether these different contingencies vary in the level of reward satiety they produce.

This study contributes to existing research in the following ways:

a) it explores whether mere human presence can serve as a reinforcer,

b) it examines the role of human gazing as a reinforcer, and

c) it evaluates the combined effect of petting and food rewards (biscuits and chicken) with gazing as reinforcers.

Additionally, we assessed the motivational value of these reinforcers by measuring the free-ranging dogs' approach latency, the duration they spent near the experimenters, and the types of behaviours they exhibited toward them. These behavioural metrics serve as proxies for understanding the motivational value of each reward type.

**Methodology**

**Ethics Statement**

The study design did not violate the Animal Ethics regulations of the Government of India (Prevention of Cruelty to Animals Act 1960, Amendment 1982). The experimental protocol was approved by the IISER Kolkata Animal Ethics Committee, as part of a larger project sanctioned by the SERB (EMR/2016/000595).

**Subjects**

We tested a total of 150 adult free-ranging dogs (75 males and 75 females), all of whom were randomly selected from the field and unfamiliar to the experimenters. To confirm that the dogs were adults, we used physical markers: descended testes in males and darkened nipples in females. Sex was determined based on external genitalia. Only healthy dogs showing no visible signs of injury or fear were included in the study. Each of the five experimental conditions involved 30 dogs.

**Study Sites**

The study was conducted in urban and semi-urban areas of Gayeshpur (22°57'29"N 88°29'33"E), Kalyani (22°58'29"N 88°26'37"E), Kataganj (22°56'58"N 88°28'14"E), Jaguli (22°55'38"N 88°32'46"E), and Anandanagar (22°58'44"N 88°29'15"E) in the Nadia district of West Bengal, India. Figure S1 shows the location of the study sites.

**Experimenters involved**

Three female experimenters, all unfamiliar to the dogs, participated in the study:

- Experimenter 1 (E1): Responsible for delivering cues and administering the experimental conditions.

- Experimenter 2 (E2): Responsible for video recording the trials.

- Experimenter 3 (E3): Assisted in distracting non-focal dogs to ensure that each test was conducted with a single focal individual.

**Experimental Design and Procedure**

Each dog was tested under one of five predetermined experimental conditions:

1. Low food reward – glucose biscuit (a piece of Parle G© biscuit), designated as "B"

2. High food reward – meat (a piece of raw chicken, weighing approximately 8–10 g), designated as "CH"

3. Petting (6 strokes from head to neck), designated as "P"

4. Control 1 – eye contact only (no physical reward), designated as "C1"

5. Control 2 – human presence without eye contact, designated as "C2"

Food and petting conditions also included direct gaze from the experimenter. The specific condition for each dog was randomly assigned. We selected chicken and biscuits as food rewards due to their ecological relevance, as these are among the most commonly encountered food items by Indian free-ranging dogs. Since people on the streets do not spend much time interacting with free-ranging dogs, we chose 10s of interaction.

Upon encountering a dog, E1 stood approximately 4-4.5 meters away and used a positive vocal cue ("ae-ae-ae") while making eye contact (Nandi et al., 2024). This calling phase lasted up to 60 seconds, or until the dog approached within 1 body length (approximately 0.8 meters) of E1- whichever occurred first. Only dogs that approached within this distance were considered for further testing.

If the dog approached, E1 administered the randomly assigned condition for approximately 10 seconds, as follows:

- In the food condition, E1 handed over the biscuit (B) or chicken piece (CH) and maintained eye contact while the dog consumed it.

- In the petting condition (P), E1 stroked the dog six times (head to neck) while maintaining eye contact.

- In Control 1 (C1), E1 maintained eye contact for 10 seconds but gave no reward. This was referred to as the gazing-only condition.

- In Control 2 (C2), E1 stood neutrally and looked at the horizon, avoiding any eye contact with the focal dog. This was referred to as the presence-only condition.

Following the administration of the condition, E1 looked away and gazed at the horizon for 60 seconds. If the dog remained within 1 body length, the same reward was delivered again after the 60s interval. If the dog moved away, E1 repeated the calling protocol (positive vocal cue and gaze) for another maximum of 60 seconds.

Each dog could be tested using the same experimental condition up to three times, depending on their continued proximity. The session ended after the dog either participated in three trials or failed to re-approach within the designated calling period of 60s. An outline of the experimental protocol is presented in Figure 1.

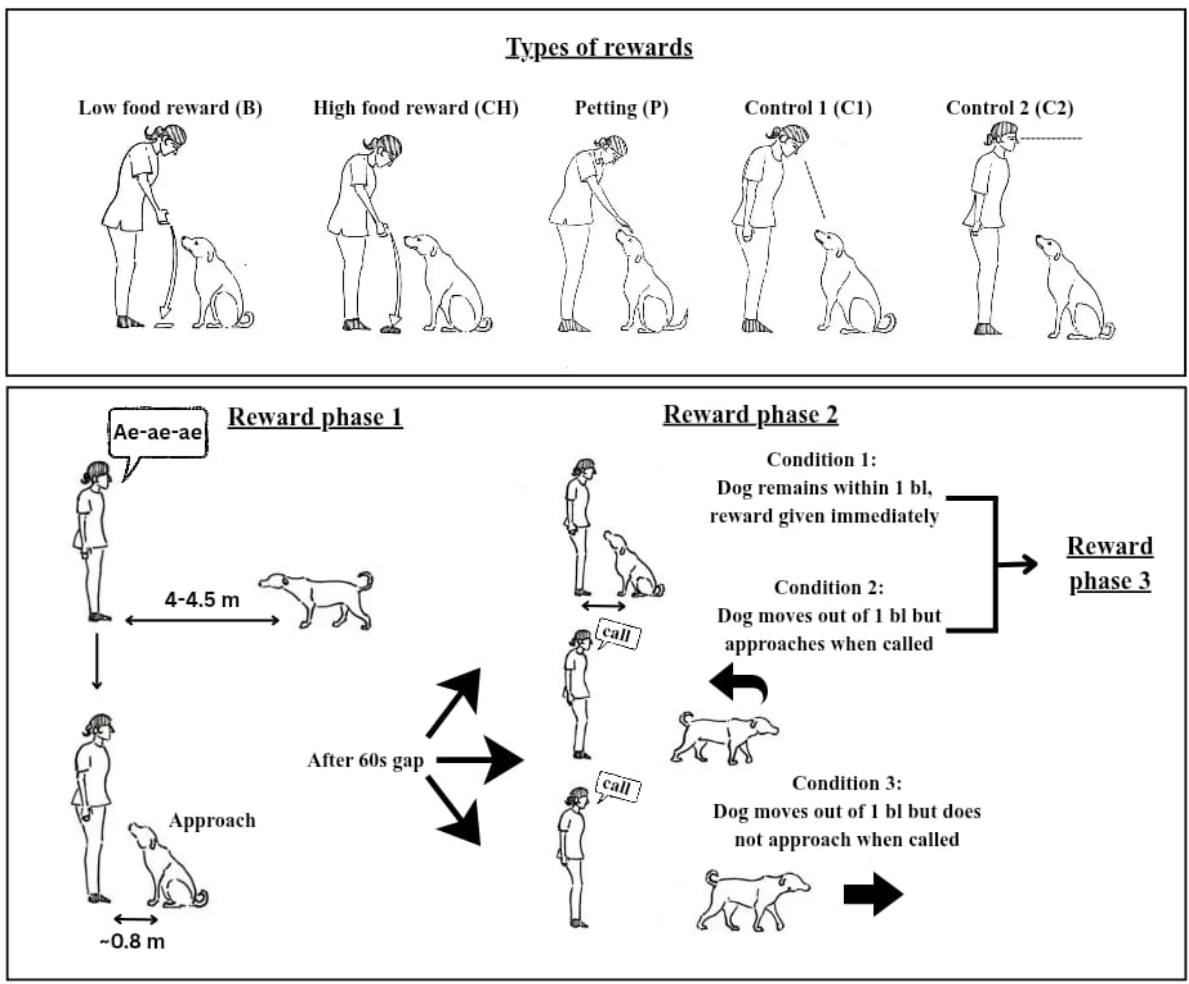

**Figure 1.** Schematic representation of the experimental protocol. The top panel illustrates the five conditions: three reward types (biscuit: B; chicken: CH; petting: P) and two controls (C1, C2). The bottom panel shows the three reward phases. If the dog approached when called the first time, it received the assigned condition; if not, the trial ended. After 60 s, reward phases 2 and 3 proceeded similarly: dogs within 1 bl received the reward, whereas dogs outside 1 bl were called once and rewarded only if they approached; otherwise, the trial was discontinued. (Illustration by Arpan Bhattacharyya).

**Video decoding and analysis**

All the videos were coded by S.N. and the obtained data was used for analysis. We measured the following parameters:

1) Number of times the reward could be provided: Each dog had the maximum probability of receiving a reward thrice. The number of times the rewards could be provided depended on whether the dogs stayed in proximity to the experimenter and/or approached when called.

2) Number of times dogs needed to be called: All the dogs had to be called for the first time. However, for the second and third times, the dogs needed to be called only if they did not stay within 1 body length of E1 for 60 seconds after the reward was given. In case the dog had moved away, E1 had to call the dog and give the reward once the dog approached within 1 body length.

3) Approach latency: The amount of time required by the dog to approach within 1 body length of E1 from the time it was called for the first time.

4) Time spent around the experimenter: The amount of time spent within 1 body length of E1 after receiving the reward. The maximum amount of possible time spent was 60s. This parameter was calculated as a proportion of the total time spent.

5) Socialization index: The Socialization Index (SI) was scored following the method described in Nandi et al. (2023), with the addition of new behaviours observed in the current study (Table 1). Only the behaviours displayed by the dogs while within 1 body length of E1 were considered for scoring SI.

**Table 1.** Behaviours exhibited by the tested dogs and their corresponding scores used to calculate the Socialization Index (SI).

| Behaviour | Score |
| --- | --- |
| Barking | -2 |
| Tail droop | -1 |
| No tail wag | 0 |
| Slow tail wag | 1 |
| Fast tail wag (without back movement) | 2 |
| Rapid tail wag (with back movement) | 3 |
| Licking experimenter | 4 |
| Pawing experimenter using forelegs/Nudging experimenter using snout/Rubbing body or head on experimenter's body | 5 |
| Jumping on experimenter's body/Belly display/Affiliative vocalization while looking at experimenter | 6 |

The number of times the reward could be provided was analysed using a cumulative link model. The response variable was modelled as an ordinal outcome with three levels using a logit link. The reference category for condition was set to "C2". The time spent near E1 was normalized by dividing the amount of time spent by the total possible time spent (60 seconds), scaling it to a range between 0 and 1. Since the number of zero and one values was minimal, a small non-zero value (0.99999 and 0.00001) was added and subtracted to ensure all transformed SI values fell strictly within this range, allowing for a beta regression. Whether the dogs needed to be called during the second and third time was analysed using a binomial logistic regression. The SI was converted into an ordinal scale. Values ≤ 0 were assigned a score of 1, values > 0 and ≤ 5 were assigned a score of 2 and values > 5 were assigned a score of 3. Thereafter, a cumulative link model was used to analyse the transformed SI scores. Approach latency was analysed using a Cox proportional hazards model. The proportional hazards assumption was met for all the models. Effect size and associated confidence intervals were mentioned for different parameters. Model-predicted estimated marginal means were also reported along with the confidence intervals.

All statistical analyses were conducted using R Studio (R Development Core Team, 2022). An alpha level of 0.05 was used for all the analyses conducted.

**Results**

**1) Did the number of times a reward could be provided vary across the reward types?**

Compared to the "C2", condition "CH" showed a statistically significant positive effect on the probability of receiving rewards ($\beta$ = 2.442, SE = 0.692, $p$ < 0.001, exp($\beta$) = 11.5, 95% CI: 2.961-44.586), suggesting that the dogs receiving chicken were 11.5 times more likely to receive a greater number of rewards compared to the ones in "C2". Condition "B" also showed a significant positive association ($\beta$ = 1.336, SE = 0.542, $p$ = 0.014, exp($\beta$) = 3.803, 95% CI: 1.315-11.0), suggesting subjects to be 3.8 times more likely to receive more biscuit rewards while conditions "P" ($\beta$ = 0.799, SE = 0.492, $p$ = 0.104) and "C1" ($\beta$ = 0.161, SE= 0.468, $p$ = 0.730) did not differ significantly from "C2". These results indicate that the condition "CH" had the strongest positive influence on reward outcomes among all tested conditions (Table S1). Figure 2 illustrates how many dogs received each reward type and how often they received it.

Pairwise comparisons of estimated marginal means revealed that several condition contrasts significantly influenced the number of rewards received. Specifically, participants in reward conditions "C2" ($\beta = -2.442$, SE = 0.692, $p = 0.004$) and "C1" ($\beta = -2.280$, SE = 0.700, $p = 0.010$) had significantly lower odds of receiving a greater number of rewards compared to condition "CH", after adjusting for multiple comparisons using Tukey's method. No other condition pairs showed statistically significant differences (Table S2).

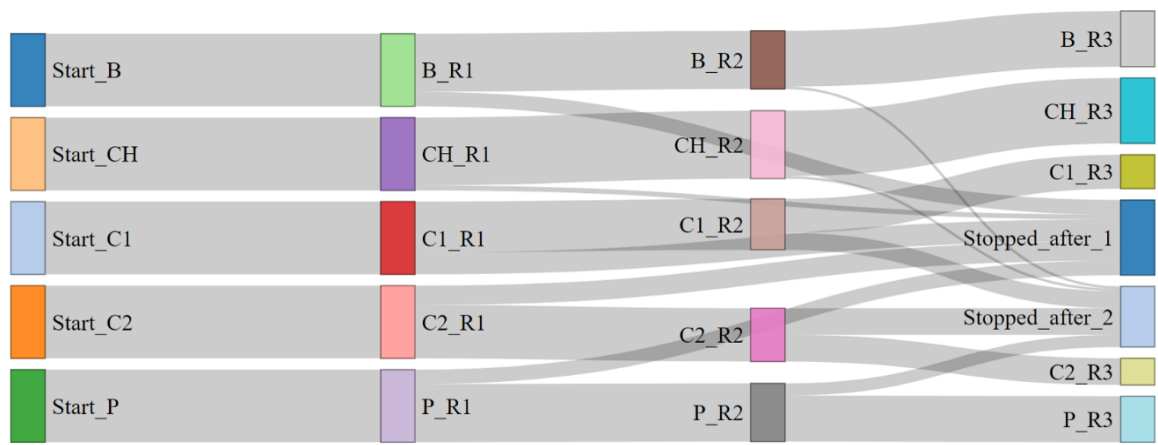

**Figure 2.** Sankey diagram illustrating the progression of dogs across reward stages (1st, 2nd, and 3rd) within five experimental conditions (B, C1, C2, CH, and P). Each condition began with 30 dogs. All dogs received at least one reward of any particular type. Flows represent the number of dogs that proceeded to subsequent rewards or stopped after a particular stage. Nodes labelled as "Start_" represent the initial allocation to each condition. Grey lines indicate transitions between reward stages (e.g., "B_R1" to "B_R2") or dropouts (e.g., "Stopped_after_1"). The thickness of the links is proportional to the number of dogs progressing or stopping at each stage.

2) **Does the amount of time spent around experimenters vary based on reward type?**

a) **Time spent after the reward was given for the first time (time1)**

Compared to the reference condition "C2", participants in condition "B" ($\beta$ = 1.603, SE = 0.345, $p$ < 0.001, exp($\beta$) = 4.971, 95% CI: 2.528 - 9.765) and condition "CH" ($\beta$ = 1.693, SE = 0.344, $p$ < 0.001, exp($\beta$) = 5.436, 95% CI: 2.77 - 10.66) spent a significantly greater proportion of time near the experimenter. However, condition "P" ($\beta$ = 0.691, SE = 0.351, $p$ = 0.049) and condition "C1" ($\beta$ = –0.011, SE = 0.349, $p$ = 0.976) did not show any difference compared to "C2" (Table S3; Figure 3a).

In a series of pairwise comparisons using odds ratios, several significant differences were observed. Compared to condition "B", condition "C2" had significantly lower odds (OR = 0.201, SE = 0.069, $p$ < 0.0001), as did "C2" compared to "CH" (OR = 0.184, SE = 0.063, $p$ < 0.0001). Similarly, "B" showed significantly higher odds than "C1" (OR = 5.020, SE = 1.730, $p$ < 0.0001) and "P" (OR = 2.488, SE = 0.822, $p$ = 0.046). Additionally, "C1" had significantly lower odds than "CH" (OR = 0.182, SE = 0.063, $p$ < 0.0001), and "CH" had higher odds than "P" (OR = 2.725, SE = 0.897, $p$ = 0.020). No significant differences were found in the contrasts between "C2" and "C1", "B" and "CH", or "C1" and "P" (all $p$ > 0.05).

### b) Time spent after the reward was given for the second time (time2)

Compared to the reference condition "C2", conditions "B" ($\beta$ = 1.320, SE = 0.389, $p$ < 0.001, exp($\beta$) = 3.745, 95% CI: 1.75 – 8.01) and "CH" ($\beta$ = 1.361, SE = 0.377, $p$ < 0.001, exp($\beta$) = 3.900, 95% CI: 1.86 – 8.16) were associated with significantly greater time spent around E1. There were no significant differences between condition "C1" ($\beta$ = 0.414, SE = 0.414, $p$ = 0.318) or condition "P" ($\beta$ = 0.446, SE = 0.400, $p$ = 0.266) compared to "C2". These findings suggest that participants in conditions "B" and "CH" spent a significantly higher amount of time after the reward was given for the second time compared to condition "C2" (Table S5; Figure 3b).

A Tukey-adjusted pairwise comparison of estimated marginal means revealed several significant differences in odds ratios between conditions. Specifically, the odds of the outcome were significantly lower in the "C2" condition compared to both "B" (OR = 0.267, SE = 0.104, $p$ = 0.006) and "CH" (OR = 0.256, SE = 0.097, *p* = 0.003), indicating that participants in the "C2" condition were less likely to spend time around E1 than those in "B" and "CH". No other contrasts reached statistical significance after adjustment (all *p* > 0.05; Table S6).

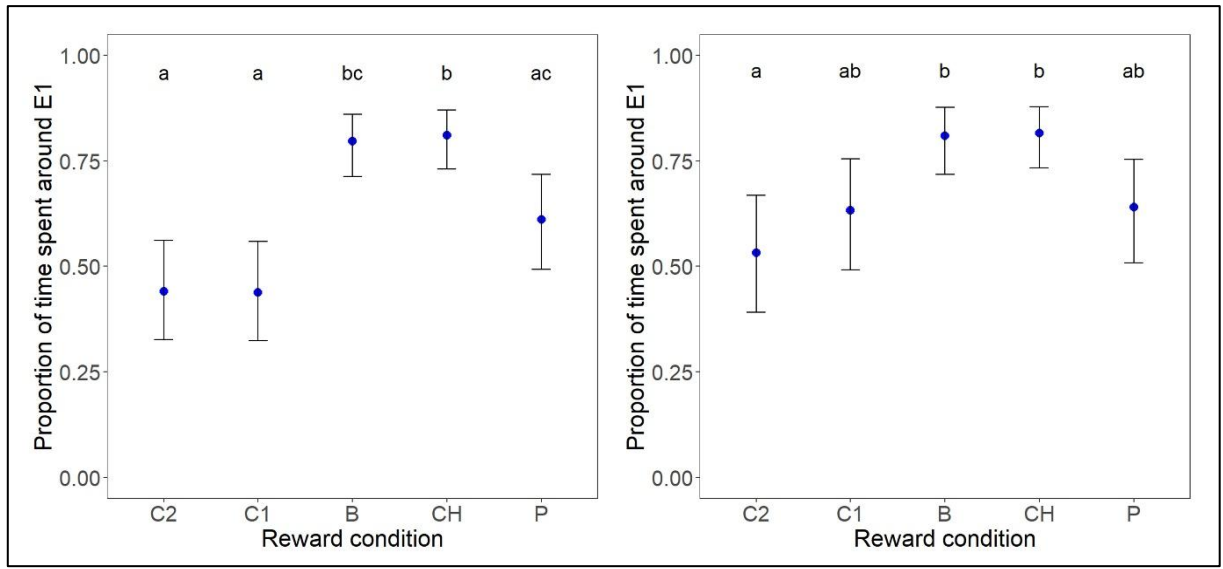

**Figure 3.** Effect plot showing the a) impact of reward condition on the amount of time spent within 1 body length of E1 when the reward was given for the first time (time 1), b) impact of reward condition on the amount of time spent within 1 body length of E1 when the reward was given for the second time (time 2). The dots represent the model-fitted mean while the whiskers represent the uncertainty (95% CI). The alphabets represent the significance levels.

### 3) Do dogs need to be called equally across reward conditions?

#### a) Likelihood of a second call (call 2)

The overall model was statistically significant ($\chi^2(4) = 30.47$, *p* < 0.001), indicating that the reward condition had a significant effect on the probability of calling. Using condition "C2" as the reference group, the results showed that dogs in the "B" condition had to be called significantly less compared to "C2" ($\beta = -1.240$, SE = 0.542, *p* = 0.022, exp($\beta$) = 0.289, 95% CI: 0.10 – 0.84), as were those in the "CH" condition ($\beta = -1.705$, SE = 0.567, *p* = 0.002, exp($\beta$) = 0.182, 95% CI: 0.060 – 0.552). No significant differences were found for the "C1" (β = 0.916, SE = 0.624, $p$ = 0.142) or "P" (β = 0.318, SE = 0.566, $p$ = 0.574) conditions (Table S7; Figure 4).

A Tukey-adjusted pairwise comparison of estimated marginal means revealed several significant differences in odds ratios between conditions. Participants in condition "C2" had significantly higher odds of being called compared to "CH" (OR = 5.500, SE = 3.110, $p$ = 0.022), and participants in "C1" also had significantly higher odds than those in "CH" (OR = 13.750, SE = 8.810, $p$ < 0.001). Additionally, participants in "B" had significantly lower odds than those in "C1" (OR = 0.116, $p$ = 0.004), and those in "B" had significantly lower odds than "P" (OR = 0.211, $p$ = 0.043). Lastly, "CH" had significantly lower odds than "P" (OR = 0.132, $p$ = 0.005). No other contrasts reached statistical significance (all $p$ > 0.05; Table S8).

### b) Likelihood of a third call (call 3)

The overall model was statistically significant ($\chi^2(4)$ = 32.01, $p$ < 0.001), indicating that condition had a significant effect on the likelihood of a third call. Compared to the reference group "C2", dogs in the "B" condition needed to be called significantly less during the third time (β = -2.097, $SE$ = 0.680, $p$ = 0.002), as were dogs in the "CH" condition (β = -2.554, $SE$ = 0.708, $p$ < 0.001). "C1" and "P" conditions did not significantly differ from the reference group (all $p$ > 0.05; Table S9; Figure 4).

A Tukey-adjusted pairwise comparison of estimated marginal means revealed several significant differences in odds ratios between conditions. Participants in "C2" had significantly higher odds of the outcome compared to those in "B" (OR = 8.143, SE = 5.540, $p$ = 0.017) and "CH" (OR = 12.857, SE = 9.100, $p$ = 0.003). Additionally, participants in "C1" had significantly higher odds than those in "CH" (OR = 9.750, SE = 6.850, $p$ = 0.011), and participants in "P" had significantly higher odds than those in "B" (OR = 0.108, SE = 0.073, $p$ = 0.009). There was also a significant difference between "CH" and "P", with CH showing lower odds (OR = 0.069, SE = 0.048, $p$ = 0.001). No other pairwise comparisons were statistically significant (all $p > 0.05$; Table S10).

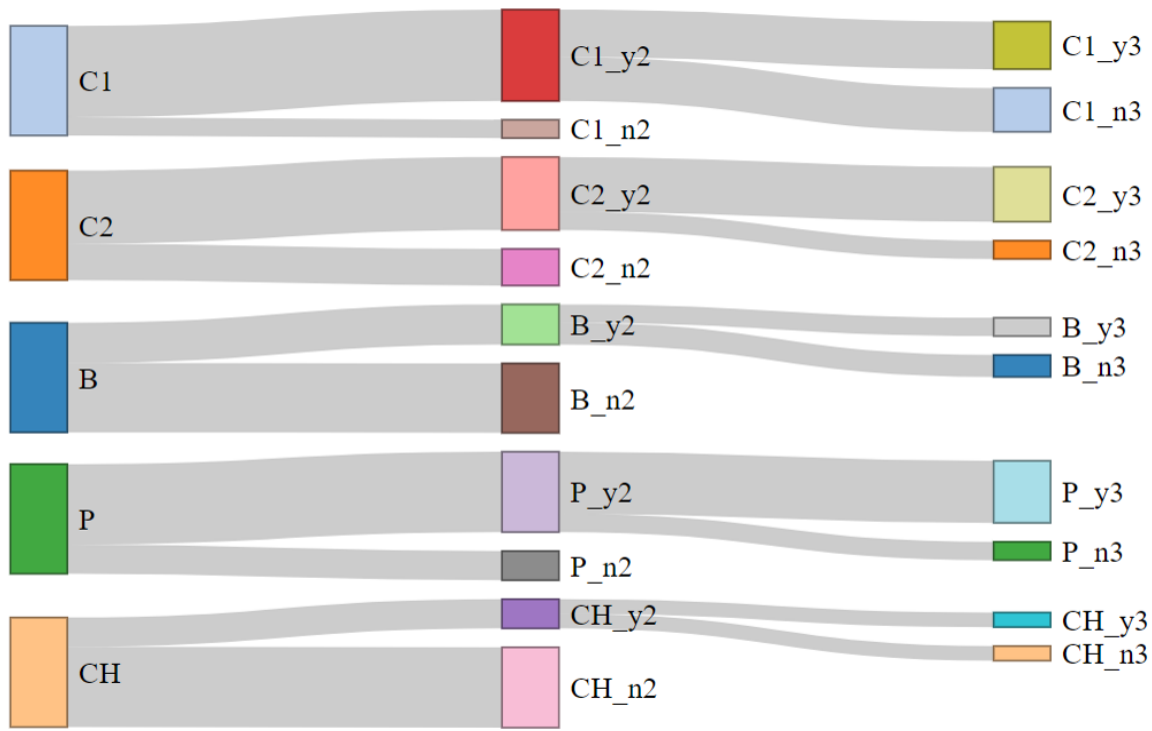

**Figure 4.** Sankey diagram illustrating the progression of dogs across the different stages of calling (1st, 2nd, and 3rd) within five experimental conditions (B, C1, C2, CH, and P). Each condition began with 30 dogs. All dogs had to be called at least once. Flows represent how many dogs proceeded to subsequent calls or stopped after a particular stage. Nodes represent the initial allocation to each condition. Grey lines indicate transitions between calls (e.g., "B_y2" indicating dogs in "B" condition that had to be called for the second time and "B_n2" indicating dogs that did not require calling). The thickness of the links is proportional to the number of dogs that were present in each stage.

4) **Does the SI vary across reward conditions?**

a) **SI during first reward provisioning (SI_1)**

The model was statistically significant for condition "P" ($\beta = 1.902$, SE = 0.586, $p = 0.001$, exp($\beta$) = 6.703, 95% CI: 2.126 – 21.126), indicating that participants in condition "P" had significantly higher odds of reporting higher behaviour scores compared to the reference group "C2". Other conditions were not statistically significant (all *p* > 0.05; Table S11), suggesting no clear evidence of a difference in behaviour scores relative to the reference condition.

A Tukey-adjusted pairwise comparison of estimated marginal means revealed a few significant differences between conditions. Participants in condition "C2" had significantly lower odds of the SI_1 compared to "P" ($\beta$ = –1.902, SE = 0.586, *p* = 0.010) and "B" had significantly lower odds compared to "P" ($\beta$ = –1.577, SE = 0.557, *p* = 0.037). No other pairwise contrasts reached statistical significance (all *p* > 0.05; Table S12; Figure 5a).

### b) SI during second reward provisioning (SI_2)

The model revealed a statistically significant effect for condition "P" (β = 1.572, SE = 0.612, *p* = 0.010), indicating that participants in condition "P" had significantly higher odds of reporting higher behaviour scores compared to the reference group "C2". Conditions "C1", "B" and "CH" were not significant (all *p* > 0.05; Table S13).

A Tukey-adjusted pairwise comparison of estimated marginal means revealed significant difference between "B" and "P" ($\beta$ = –1.626, SE = 0.582, *p* = 0.042), indicating dogs in "B" to display significantly lower SI_2 compared to "P". No other pairwise contrasts reached statistical significance (all *p* > 0.05; Table S14; Figure 5b).

### c) SI during third reward provisioning (SI_3)

The model showed no statistically significant effects for any condition relative to the reference group (all *p*> 0.05; Table S15; Figure 5c), indicating no clear evidence that reward condition influenced SI_3.

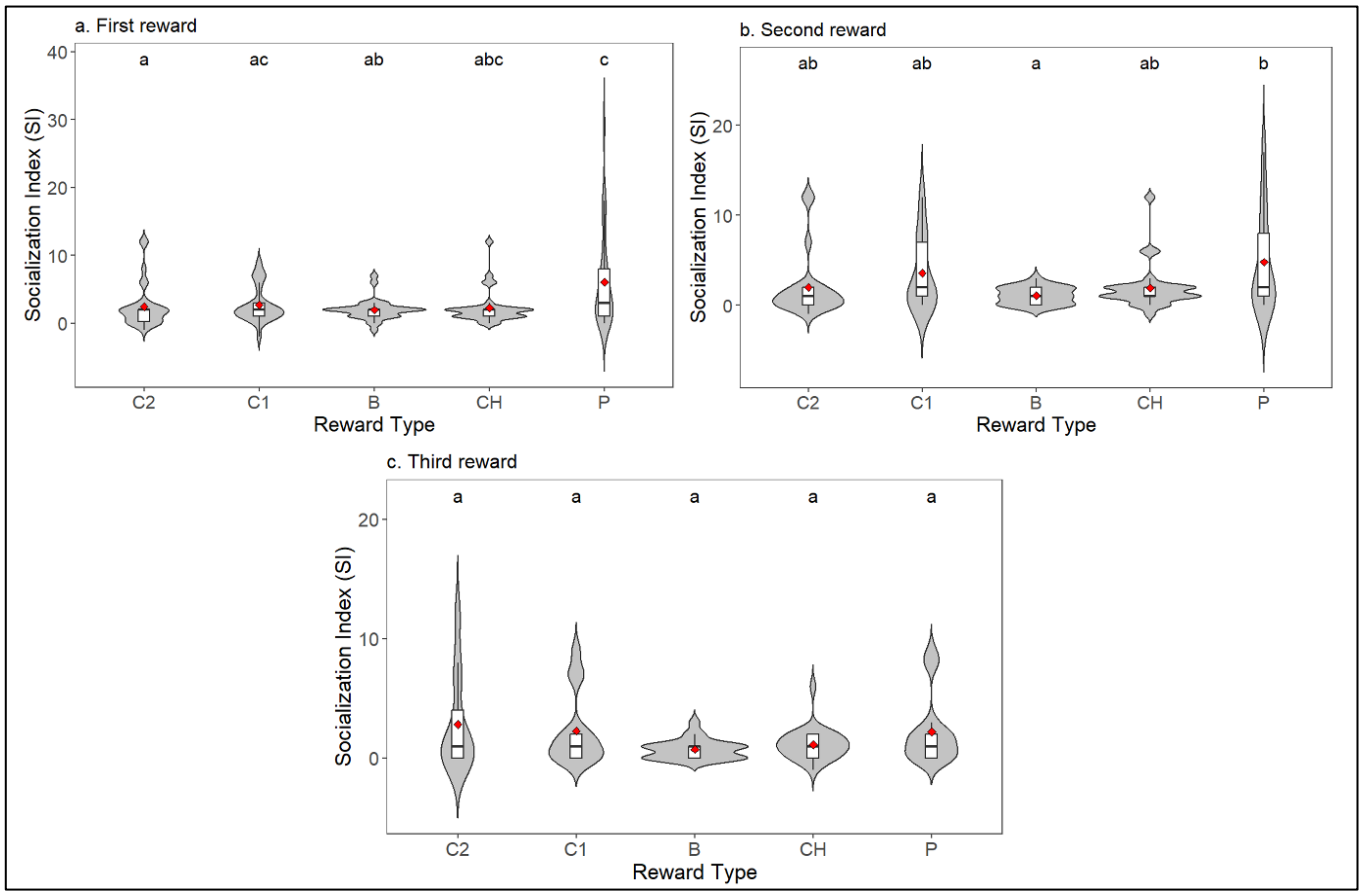

**Figure 5.** Violin plots with overlaid boxplots showing SI scores towards E1 across different reward types: (a) after the first reward (SI_1), (b) after the second reward (SI_2), and (c) after the third reward (SI_3). The red diamond indicates the mean, the box denotes the interquartile range, and the black line within the box marks the median. Letters indicate significant differences.

d) Does the latency of approach vary across rewards?

Compared to "C2", none of the reward conditions showed a significant difference in approach latency during the first and second approach (Table S16, Table S17; Figure 6a, 6b). However, compared to "C2", dogs in the "B" ($HR = 5.106$, 95% CI: 1.258 - 20.724, $p = 0.022$), "CH" ($HR = 5.725$, 95% CI: 1.270 - 25.804, $p = 0.023$), and "P" condition ($HR = 3.937$, 95% CI: 1.262 - 12.274, $p = 0.018$) had significantly higher hazard rates of approaching during the third time, indicating shorter approach latencies. The "C1" condition did not significantly differ from the reference group ($HR = 1.715$, 95% CI: 0.484 - 6.081, $p = 0.404$; Table S18; Figure 6c).

A Tukey-adjusted pairwise comparison of estimated marginal means, however, revealed no significant differences in approach latency during the third time across reward types (all p>0.05; Table S19).

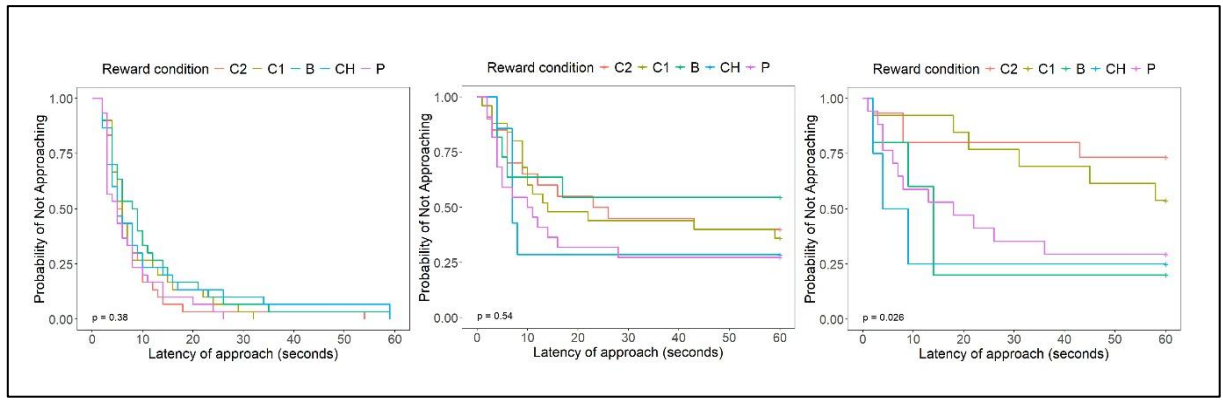

**Figure 6.** Kaplan–Meier survival curves illustrating the time to approach (seconds) in dogs receiving different reward types. a) when the dog was called for the first time, b) when the dog was called for the second time, and c) when the dog was called for the third time. Statistical comparison across days was performed using the log-rank test.

**Discussion**

The present study explored how free-ranging dogs respond to different reward types offered by unfamiliar humans. By systematically varying the nature of the reward, ranging from high-value food (chicken), low-value food (biscuits), social interaction (petting), to control conditions (presence-only and gazing only), we examined how these dogs modulate their behaviour in response to varying incentives. Our findings reveal distinct patterns in motivation, social responsiveness, and proximity behaviour across reward conditions, offering valuable insights into how free-ranging dogs assess and interact with unfamiliar humans in urban settings.

Dogs offered chicken were significantly more likely to receive a higher number of rewards compared to those in the control conditions (presence-only and gazing-only). This finding confirms that high-value food is a potent driver of approach behaviour in free-ranging dogs. In contrast, biscuits and petting did not lead to a significantly higher number of rewards than the control conditions. Furthermore, the number of rewards received in these two groups did not differ statistically from that of the chicken group. Together, we see a pattern in the results that strongly suggest a hierarchy of reward value (Chicken > Biscuit/Petting > Controls) where our study was only statistically powered to detect the largest difference, that between the highest-value reward and the controls. This shows the lower but significant importance of low-value food (biscuit) and petting as rewards compared to the chicken condition. The non-significant findings for biscuits and petting are likely due to their intermediate motivational value, which was not high enough to be statistically distinguished from either the controls or the high-value chicken reward. Previous work has shown a preference for chicken over biscuits among Indian free-ranging dogs (Bhadra & Bhadra, 2014). Further, chicken rewards elicit higher and steeper levels of conflict between free-ranging dog mothers and their pups during weaning, as compared to biscuits (Paul et al., 2015). This study provides another context in support of the relative importance of the two kinds of food rewards for the dogs. We show that in a free-choice context, a low-value reward like a biscuit may be insufficient to significantly motivate approach, even in an environment where food is typically scarce. Although the hunger levels of the randomly selected dogs could not be controlled, the clear dependence on reward quality suggests that motivation was not determined by hunger alone. Finally, our results are however, not in line with the findings of Feuerbacher et al. (2012), who found lower satiation in dogs receiving food compared to petting.

The time spent by the dogs near the experimenter after receiving a reward also mirrored this trend. Dogs receiving chicken and biscuits stayed significantly longer after both the first and second rewards compared to dogs receiving only human presence. This suggests that although biscuits may not universally motivate approach, they can be effective for some dogs, perhaps those already more inclined towards human interaction. Dogs in the petting condition, by contrast, spent less time after the first reward compared to the chicken group, but these differences disappeared after the second reward. This may indicate that social rewards like petting require more time to build trust or familiarity. This finding aligns with previous research showing that while food is preferred during initial interactions, tactile social rewards can become more or at least equally effective with repeated contact (Bhattacharjee et al., 2021; Nandi et al., 2025, *under review*). Moreover, dogs receiving either chicken or biscuits spent more time near the experimenter after the first reward compared to dogs in the gazing condition. However, these differences were not maintained after the second reward, suggesting that human gaze may itself act as a subtle positive reinforcement. Dogs might interpret direct gazing as a form of attention or anticipation of a reward, consistent with earlier findings showing increased begging when humans gaze at dogs while eating (Biswas et al., 2025; *preprint*). Interestingly, the amount of time spent in the two control conditions (only gazing and only human presence) remained similar after the rewards were given. This suggests that human presence alone generates motivation in dogs, likely a form of reward anticipation, at levels comparable to being gazed at. This, however, contradicts earlier findings (Biswas et al., 2025; *preprint*) where dogs showed greater begging propensity when the person eating gazed at them. This difference might be explained by the fact that our study did not involve food being present in the control conditions, creating uncertainty regarding the expectation of a reward. On the other hand, the other study involved the dog watching the experimenter eat, which provided a higher level of motivation and caused them to beg more in response to the gaze.

One particularly compelling result was the reduced need for repeated calls in the chicken condition. Dogs receiving chicken were significantly less likely to require a second or third call to approach the experimenter compared to the dogs receiving either gaze-only or presence-only conditions, indicating a stronger and more immediate motivation. Conversely, dogs in the petting condition required more repeated calling during the second and third time compared to dogs receiving chicken and biscuits, pointing to a lower reward value compared to food-based conditions. This distinction highlights how different reward types produce varying levels of satiety or engagement, with food eliciting the higher behavioural responses.

The Socialization Index (SI) scores added nuance to these results. While dogs in the petting condition showed higher SI scores after the first and second rewards, this pattern declined by the third provisioning, possibly due to a habituation effect. A similar effect has been observed among free-ranging dogs in an earlier study (Nandi et al., 2025; *under review*). Interestingly, while petting led to higher SI scores than dogs receiving biscuits, it did not elicit significantly higher scores than dogs receiving chicken. This suggests that although petting fosters affiliative behaviour, it may not be more rewarding than high-value food in these contexts. Also, the SI scores displayed by the dogs receiving petting were not found to be higher than dogs receiving only human gazing, again pointing to the importance of human gaze as a positive reinforcer.

Approach latency provided further insights. It was not found to differ after the first, second or third call across any of the reward types. This shows that although the reward type affected their decision to approach, the time spent in proximity to the reward giver and the behaviours displayed, it did not affect the time taken to approach.

Although food proved to be the most effective motivator overall, the petting condition yielded notable insights. The elevated Socialization Index (SI) scores suggest that tactile interaction acts as a meaningful reward, despite eliciting more rapid satiation compared to high-value food. This implies that social rewards may be effective primarily for a specific subset of dogs, likely those with higher baseline sociability or a history of positive human interaction during their ontogeny. While such developmental data were unavailable in the current study, these findings underscore the need for future research into how life history and personality shape reward preferences in free-ranging populations. Also, it would be quite interesting to check if increasing the time of social interaction can result in promoting similar behavioural responses in dogs receiving petting and food, as was observed by Fonberg et al. (1981).

Altogether, our results underscore the behavioural flexibility of free-ranging dogs, who are capable of adjusting their approach, proximity, and social behaviour based on the perceived value of rewards offered by unfamiliar humans. Placed within the broader context of urban adaptation, these findings highlight the sophisticated decision-making strategies that free-ranging dogs utilise to survive in a human-dominated ecological niche. The clear preference for high-value food over social contact with strangers reflects an optimal foraging strategy (Cowie, 1977; Pyke, 1984) that balances energy intake against the potential risks of interacting with unfamiliar humans. This also suggests that such high-value resources can act as attractants for these dogs in certain microhabitats that provide ample opportunities to explore such resources, like dumping sites, meat shops or sites where human feeders provide meat scraps regularly. Ultimately, the ability of these dogs to read human cues and weigh reward values demonstrates a high degree of adaptability that is central to their persistence in the urban landscape, where human behaviour can range from rewarding to threatening.

**Acknowledgements**

The authors are thankful to Ms. Krutika Gajanan Umare for her contribution to fieldwork and Arpan Bhattacharyya for making the illustration for the experimental protocol.

**Availability of data and materials**

Supplementary data associated with this article can be found in the online version at doi: 10.17632/2y96sxychj.1

**Funding**

This research was supported by the Department of Science and Technology, Ministry of Science and Technology (INSPIRE fellowship to S.N.) and the Janaki Ammal – National Women Bioscientist Award (BT/HRD/NBA-NWB/39/2020-21 (YC-1)), Department of Biotechnology, India.

Zhang, Z., Khederzadeh, S., & Li, Y. (2020). Deciphering the puzzles of dog domestication. *Zoological research*, *41*(2), 97.

**CRediT authorship contribution statement**

**Srijaya Nandi:** Conceptualization, Methodology, Formal analysis, Investigation, Data curation, Writing – Original Draft, Visualization, Project administration. **Dipayanti Aditya:** Investigation. **Tithi Chakraborty:** Investigation. **Rachael Sara Paul:** Investigation. **Anindita Bhadra:** Conceptualization, Methodology, Resources, Writing – Review & Editing, Supervision, Funding acquisition.

**Conflict of interest**

The authors declare no conflict of interest.

**Supplementary material**

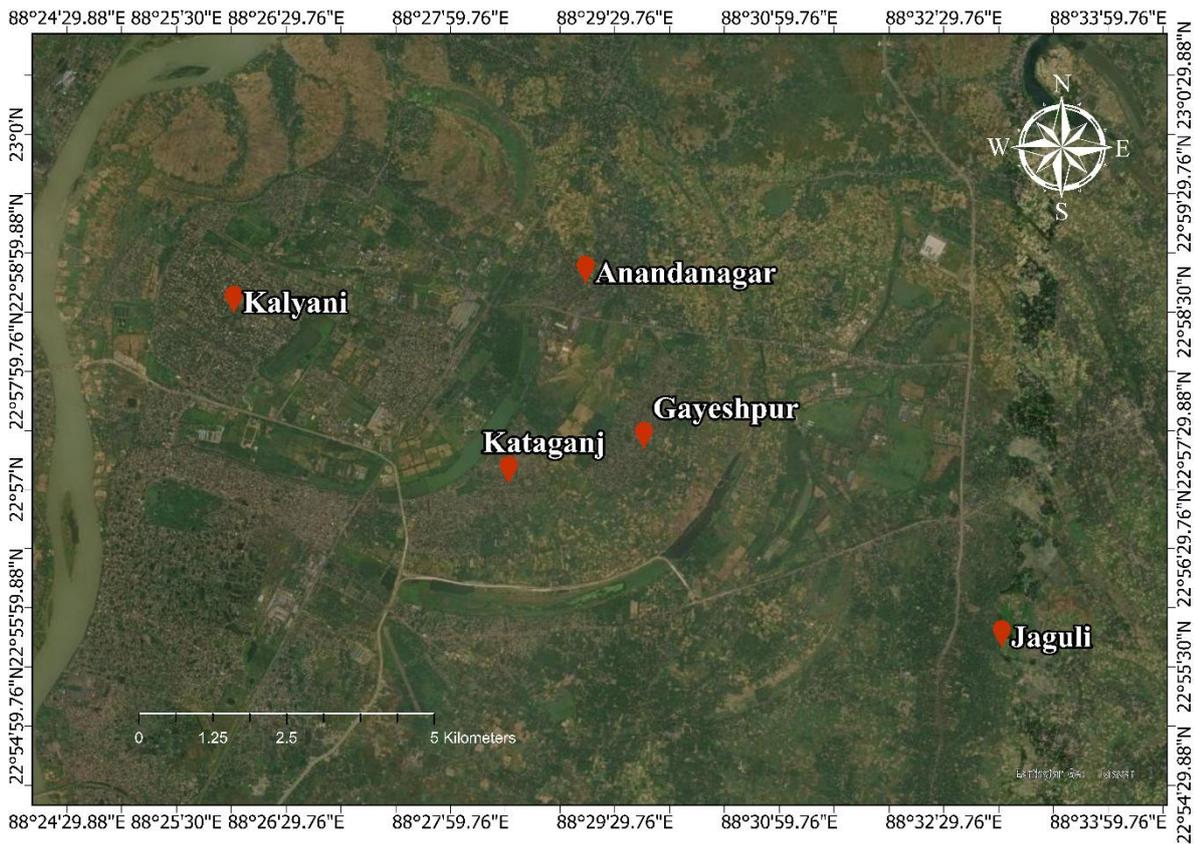

**Figure S1.** Study sites

**Table S1.** Results of the cumulative link model to assess the influence of reward condition on the number of times a reward could be provided.

|              | Estimate | Std. Error | z value | Pr (>|z|) |
| ------------ | -------- | ---------- | ------- | --------- |
| conditionB   | 1.336    | 0.542      | 2.464   | 0.014     |
| conditionC1  | 0.161    | 0.468      | 0.344   | 0.730     |
| conditionCH  | 2.442    | 0.692      | 3.531   | <0.001    |
| conditionP   | 0.799    | 0.492      | 1.625   | 0.104     |

**Table S2.** Pairwise Comparisons of Estimated Marginal Means to assess the effect of reward condition on number of times a reward could be provided (Tukey-Adjusted p-values)

| Contrast | Estimate | SE | z.ratio | p.value |
| -------- | -------- | -- | ------- | ------- |
|          |          |    |         |         |
| C2 - B | -1.336 | 0.542 | -2.464 | 0.099 |
| C2 - C1 | -0.161 | 0.468 | -0.344 | 0.997 |
| C2 - CH | -2.442 | 0.692 | -3.531 | 0.004 |
| C2 - P | -0.799 | 0.492 | -1.625 | 0.481 |
| B - C1 | 1.175 | 0.553 | 2.215 | 0.209 |
| B - CH | -1.106 | 0.747 | -1.469 | 0.576 |
| B - P | 0.537 | 0.570 | 0.942 | 0.880 |
| C1 - CH | -2.280 | 0.700 | -3.259 | 0.010 |
| C1 - P | -0.638 | 0.504 | -1.266 | 0.712 |
| CH - P | 1.643 | 0.713 | 2.304 | 0.144 |

**Table S3.** Results of a beta regression model predicting the effect of reward condition on "time 1".

| | Estimate | Std. Error | z value | Pr (>|z|) |
|---|---|---|---|---|
| (Intercept) | -0.239 | 0.248 | -0.967 | 0.334 |
| conditionB | 1.603 | 0.345 | 4.647 | <0.001 |
| ConditionC1 | -0.011 | 0.349 | -0.031 | 0.976 |
| ConditionCH | 1.693 | 0.344 | 4.923 | <0.001 |
| ConditionP | 0.691 | 0.351 | 1.968 | 0.049 |

**Table S4.** Pairwise Comparisons of Odds Ratios Between Reward Conditions for "time 1"

| contrast | odds.ratio | SE | z.ratio | p.value |
|---|---|---|---|---|
| C2 / B | 0.201 | 0.069 | -4.647 | <0.0001 |
| C2 / C1 | 1.011 | 0.353 | 0.031 | 1.000 |
| C2 / CH | 0.184 | 0.063 | -4.923 | <0.0001 |
| C2 / P | 0.501 | 0.176 | -1.968 | 0.282 |
| B / C1 | 5.020 | 1.730 | 4.676 | <0.0001 |
| B / CH | 0.913 | 0.282 | -0.294 | 0.998 |
| B / P | 2.488 | 0.822 | 2.758 | 0.046 |
| C1 / CH | 0.182 | 0.063 | -4.952 | <0.0001 |
| C1 / P | 0.496 | 0.174 | -1.998 | 0.267 |
| CH / P | 2.725 | 0.897 | 3.045 | 0.020 |

**Table S5.** Results of a beta regression model predicting the effect of reward condition on "time 2".

| | Estimate | Std. Error | z value | Pr (>|z|) |
|---|---|---|---|---|
| (Intercept) | 0.129 | 0.292 | 0.442 | 0.659 |
| conditionB | 1.320 | 0.389 | 0.393 | <0.001 |
| ConditionC1 | 0.414 | 0.414 | 1.000 | 0.318 |
| ConditionCH | 1.361 | 0.377 | 3.607 | <0.001 |
| ConditionP | 0.446 | 0.400 | 1.113 | 0.266 |

**Table S6.** Pairwise Comparisons of Odds Ratios Between Reward Conditions for "time 2"

| contrast | odds.ratio | SE | z.ratio | p.value |
|---|---|---|---|---|
| C2 / B | 0.267 | 0.104 | -3.393 | 0.006 |
| C2 / C1 | 0.661 | 0.274 | -1.000 | 0.856 |
| C2 / CH | 0.256 | 0.097 | -3.607 | 0.003 |
| C2 / P | 0.640 | 0.256 | -1.113 | 0.800 |

| contrast | odds.ratio | SE | z.ratio | p.value |
|---|---|---|---|---|
| B / C1 | 2.474 | 0.945 | 2.372 | 0.123 |
| B / CH | 0.959 | 0.317 | -0.126 | 1.000 |
| B / P | 2.396 | 0.879 | 2.382 | 0.120 |
| C1 / CH | 0.388 | 0.143 | -2.560 | 0.078 |
| C1 / P | 0.968 | 0.385 | -0.081 | 1.000 |
| CH / P | 2.498 | 0.885 | 2.582 | 0.074 |

**Table S7.** Results of a binomial logistic regression model predicting the effect of reward condition on the likelihood of second call (call 2)

|  | Estimate | Std. Error | z value | Pr (>|z|) |
|---|---|---|---|---|
| (Intercept) | 0.693 | 0.387 | 1.790 | 0.074 |
| conditionB | -1.240 | 0.542 | -2.288 | 0.022 |
| ConditionC1 | 0.916 | 0.624 | 1.467 | 0.142 |
| ConditionCH | -1.705 | 0.566 | -3.011 | 0.003 |
| ConditionP | 0.318 | 0.566 | 0.563 | 0.534 |

**Table S8.** Pairwise Comparisons of Odds Ratios Between Reward Conditions for "call 2"

| contrast | odds.ratio | SE | z.ratio | p.value |
|---|---|---|---|---|
| C2 / B | 3.455 | 1.870 | 2.288 | 0.149 |
| C2 / C1 | 0.400 | 0.250 | -1.467 | 0.584 |
| C2 / CH | 5.500 | 3.110 | 3.011 | 0.022 |
| C2 / P | 0.727 | 0.412 | -0.563 | 0.980 |
| B / C1 | 0.116 | 0.072 | -3.481 | 0.004 |
| B / CH | 1.592 | 0.892 | 0.830 | 0.921 |
| B / P | 0.211 | 0.118 | -2.781 | 0.043 |
| C1 / CH | 13.750 | 8.810 | 4.091 | <0.001 |
| C1 / P | 1.818 | 1.160 | 0.933 | 0.884 |
| CH / P | 0.132 | 0.077 | -3.465 | 0.005 |

**Table S9.** Results of a binomial logistic regression model predicting the effect of reward condition on the likelihood of third call (call 3)

| | Estimate | Std. Error | z value | Pr (>|z|) |
|---|---|---|---|---|
| (Intercept) | 0.762 | 0.458 | 1.665 | 0.096 |
| conditionB | -2.097 | 0.680 | -3.085 | 0.002 |
| ConditionC1 | -0.277 | 0.641 | -0.431 | 0.666 |
| ConditionCH | -2.554 | 0.708 | -3.607 | <0.001 |
| ConditionP | 0.125 | 0.641 | 0.195 | 0.845 |

**Table S10.** Pairwise Comparisons of Odds Ratios Between Reward Conditions for "call 3"

| contrast | odds.ratio | SE | z.ratio | p.value |
|---|---|---|---|---|
| C2 / B | 8.143 | 5.540 | 3.085 | 0.017 |
| C2 / C1 | 1.319 | 0.846 | 0.431 | 0.993 |
| C2 / CH | 12.857 | 9.100 | 3.607 | 0.003 |
| C2 / P | 0.882 | 0.566 | -0.195 | 1.000 |
| B / C1 | 0.162 | 0.109 | -2.700 | 0.054 |
| B / CH | 1.579 | 1.160 | 0.619 | 0.972 |
| B / P | 0.108 | 0.073 | -3.297 | 0.009 |
| C1 / CH | 9.750 | 6.850 | 3.241 | 0.010 |
| C1 / P | 0.669 | 0.425 | -0.632 | 0.970 |
| CH / P | 0.069 | 0.048 | -3.814 | 0.001 |

Table S11. Results of the cumulative link model to assess the influence of reward condition on the SI after first reward provisioning (SI_1)

| | Estimate | Std. Error | z value | Pr (>\|z\|) |
|---|---|---|---|---|
| conditionB | 0.325 | 0.572 | 0.568 | 0.570 |
| conditionC1 | 0.880 | 0.588 | 1.498 | 0.134 |
| conditionCH | 0.595 | 0.581 | 1.025 | 0.306 |
| conditionP | 1.902 | 0.586 | 3.244 | 0.001 |

Table S12. Pairwise Comparisons of Estimated Marginal Means to assess the effect of reward condition on SI_1 (Tukey-Adjusted p-values)

| Contrast | Estimate | SE | z.ratio | p.value |
|---|---|---|---|---|
| C2 - B | -0.325 | 0.572 | -0.568 | 0.980 |
| C2 - C1 | -0.880 | 0.588 | -1.498 | 0.564 |
| C2 - CH | -0.595 | 0.581 | -1.025 | 0.844 |
| C2 - P | -1.902 | 0.586 | -3.244 | 0.010 |
| B - C1 | -0.556 | 0.564 | -0.984 | 0.862 |
| B - CH | -0.270 | 0.560 | -0.483 | 0.989 |
| B - P | -1.577 | 0.557 | -2.833 | 0.037 |
| C1 - CH | 0.285 | 0.566 | 0.504 | 0.987 |
| C1 - P | -1.021 | 0.545 | -1.873 | 0.332 |
| CH - P | -1.307 | 0.552 | -2.366 | 0.125 |

**Table S13.** Results of the cumulative link model to assess the influence of reward condition on the SI after second reward provisioning (SI_2)

| | Estimate | Std. Error | z value | Pr (>|z|) |
|---|---|---|---|---|
| conditionB | -0.054 | 0.566 | -0.096 | 0.924 |
| conditionC1 | 1.213 | 0.632 | 1.919 | 0.055 |
| conditionCH | 0.680 | 0.558 | 1.218 | 0.223 |
| conditionP | 1.572 | 0.612 | 2.570 | 0.010 |

**Table S14.** Pairwise Comparisons of Estimated Marginal Means to assess the effect of reward condition on SI_2 (Tukey-Adjusted p-values)

| Contrast | Estimate | SE | z.ratio | p.value |
|---|---|---|---|---|
| C2 - B | 0.054 | 0.566 | 0.096 | 1.000 |
| C2 - C1 | -1.213 | 0.632 | -1.919 | 0.307 |
| C2 - CH | -0.680 | 0.558 | -1.218 | 0.741 |
| C2 - P | -1.572 | 0.612 | -2.570 | 0.076 |
| B - C1 | -1.270 | 0.603 | -2.101 | 0.219 |
| B - CH | -0.734 | 0.525 | -1.399 | 0.628 |
| B - P | -1.626 | 0.582 | -2.794 | 0.042 |
| C1 - CH | 0.533 | 0.580 | 0.919 | 0.890 |

| | | | | | |
|---|---|---|---|---|---|
| C1 - P | -0.359 | 0.611 | -0.588 | 0.977 |
| CH - P | -0.892 | 0.553 | -1.612 | 0.490 |

**Table S15.** Results of the cumulative link model to assess the influence of reward condition on the SI after third reward provisioning (SI_3)

| | Estimate | Std. Error | z value | Pr (>|z|) |
|---|---|---|---|---|
| conditionB | -0.576 | 0.758 | -0.760 | 0.447 |
| conditionC1 | 0.199 | 0.848 | 0.234 | 0.815 |
| conditionCH | -0.057 | 0.741 | -0.078 | 0.938 |
| conditionP | 0.183 | 0.791 | 0.231 | 0.818 |

**Table S16.** Results of Cox Proportional Hazards Model examining the effect of reward condition on approach latency after the first call.

| | coef | exp(coef) | se(coef) | z | Pr(>|z|) |
|---|---|---|---|---|---|
| conditionB | -0.349 | 0.705 | 0.260 | -1.341 | 0.180 |
| conditionC1 | -0.129 | 0.879 | 0.260 | -0.496 | 0.620 |
| conditionCH | -0.258 | 0.773 | 0.264 | -0.977 | 0.329 |
| conditionP | 0.125 | 1.133 | 0.259 | 0.481 | 0.631 |

| | exp(coef) | exp(-coef) | lower .95 | upper .95 |
|---|---|---|---|---|
| conditionB | 0.705 | 1.418 | 0.423 | 1.175 |
| conditionC1 | 0.879 | 1.138 | 0.528 | 1.463 |
| conditionCH | 0.773 | 1.294 | 0.460 | 1.296 |
| conditionP | 1.133 | 0.883 | 0.681 | 1.883 |

Table S17. Results of Cox Proportional Hazards Model examining the effect of reward condition on approach latency after the second call.

| | coef | exp(coef) | se(coef) | z | Pr(>\|z\|) |
|---|---|---|---|---|---|
| conditionB | -0.301 | 0.740 | 0.532 | -0.565 | 0.572 |
| conditionC1 | 0.058 | 1.060 | 0.382 | 0.152 | 0.879 |
| conditionCH | 0.423 | 1.527 | 0.535 | 0.792 | 0.429 |
| conditionP | 0.434 | 1.543 | 0.382 | 1.134 | 0.257 |

| | exp(coef) | exp(-coef) | lower .95 | upper .95 |
|---|---|---|---|---|
| conditionB | 0.740 | 1.351 | 0.261 | 2.101 |
| conditionC1 | 1.060 | 0.943 | 0.501 | 2.242 |
| conditionCH | 1.527 | 0.655 | 0.535 | 4.356 |
| conditionP | 1.543 | 0.648 | 0.729 | 3.267 |

Table S18. Results of Cox Proportional Hazards Model examining the effect of reward condition on approach latency after the third call

| | coef | exp(coef) | se(coef) | z | Pr(>\|z\|) |
|---|---|---|---|---|---|
| conditionB | 1.630 | 5.106 | 0.715 | 2.281 | 0.022 |
| conditionC1 | 0.539 | 1.715 | 0.646 | 0.835 | 0.404 |
| conditionCH | 1.745 | 5.725 | 0.768 | 2.271 | 0.023 |
| conditionP | 1.370 | 3.936 | 0.580 | 2.362 | 0.182 |

|  | exp(coef) | exp(-coef) | lower .95 | upper .95 |
|---|---|---|---|---|
| conditionB | 5.106 | 0.196 | 1.258 | 20.724 |
| conditionC1 | 1.715 | 0.583 | 0.484 | 6.081 |
| conditionCH | 5.725 | 0.175 | 1.270 | 25.804 |
| conditionP | 3.937 | 0.254 | 1.262 | 12.274 |

**Table S19.** Pairwise Comparisons of Estimated Marginal Means to assess the effect of reward condition on approach latency after the third call (Tukey-Adjusted p-values)

| contrast | ratio | SE | df | null | z.ratio | p.value |
|---|---|---|---|---|---|---|
| C2/B | 0.196 | 0.140 | Inf | 1 | -2.281 | 0.151 |
| C2/C1 | 0.583 | 0.377 | Inf | 1 | -0.835 | 0.920 |
| C2/CH | 0.175 | 0.134 | Inf | 1 | -2.271 | 0.154 |
| C2/P | 0.254 | 0.147 | Inf | 1 | -2.362 | 0.126 |
| B/C1 | 2.977 | 1.940 | Inf | 1 | 1.671 | 0.452 |
| B/CH | 0.892 | 0.683 | Inf | 1 | -0.149 | 0.999 |
| B/P | 1.297 | 0.753 | Inf | 1 | 0.448 | 0.992 |
| C1/CH | 0.300 | 0.213 | Inf | 1 | -1.693 | 0.438 |
| C1/P | 0.436 | 0.219 | Inf | 1 | -1.653 | 0.463 |
| CH/P | 1.454 | 0.941 | Inf | 1 | 0.579 | 0.978 |